\documentclass[aps,pra,twocolumn,showpacs,amsmath,amssymb,nofootinbib,longbibliography
]{revtex4-1}

\usepackage[pdftex]{graphicx}
\usepackage[usenames,dvipsnames]{xcolor}
\usepackage{epstopdf}
\usepackage[hidelinks]{hyperref}

\usepackage[normalem]{ulem}

\graphicspath{{FIGs/}}

 \usepackage[utf8]{inputenc}

\usepackage{beramono}
\usepackage{listings}

\usepackage{amssymb,amsmath,amsthm}

\theoremstyle{definition}

\theoremstyle{remark}

\lstdefinelanguage{Julia}%
  {morekeywords={abstract,break,case,catch,const,continue,do,else,elseif,%
      end,export,false,for,function,immutable,import,importall,if,in,%
      macro,module,otherwise,quote,return,switch,true,try,type,typealias,%
      using,while},%
   sensitive=true,%
   alsoother={\$},%
   morecomment=[l]\#,%
   morecomment=[n]{\#=}{=\#},%
   morestring=[s]{"}{"},%
   morestring=[m]{'}{'},%
     literate={é}{{\'e}}1
           {è}{{\`e}}1
           {ù}{{\`u}}1
}[keywords,comments,strings]%

\usepackage{chngcntr}

\begin{document}

\title{Block Lanczos method for excited states on a quantum computer}
\author{Thomas E.~Baker}
\affiliation{Department of Physics, University of York, Heslington,York YO10 5DD, United Kingdom}

\begin{abstract}

The method of quantum Lanczos recursion is extended to solve for multiple excitations on the quantum computer.  While quantum Lanczos recursion is in principle capable of obtaining excitations, the extension to a block Lanczos routine can resolve degeneracies with better precision and only costs $\mathcal{O}(d^2)$ for $d$ excitations on top of the previously introduced quantum Lanczos recursion method.  The formal complexity in applying all operators to the system at once with oblivious amplitude amplification is exponential, but this cost can be kept small to obtain the ground state by incrementally adding operators. The error of the ground state energy based on the accuracy of the Lanczos coefficients is investigated and the error of the ground state energy. It is demonstrated to scale linearly with the uncertainty of the Lanczos coefficients. Extension to non-Hermitian operators is also discussed.

\end{abstract}
\maketitle


\section{Introduction}

For quantum computation to lead to new discoveries \cite{nielsen2010quantum}, efficient means of solving for the ground state must be understood and implemented. Some near term algorithms that have been used in the era of noisy quantum devices have led to an increased interest in results obtained from variational quantum eigensolvers \cite{o2016scalable,kandala2017hardware,gard2020efficient,gokhale2020n,ollitrault2020quantum,klymko2021real,kremenetski2021simulation} which ultimately face noise and other limitations \cite{bittel2021training}.  Looking to the long term capabilities of a quantum computer when error correction is available, other algorithms are useful to investigate.

By far the longest studied algorithms to obtain ground states is the implementation of real time evolution. In this algorithm, an initial Hamiltonian $\mathcal{H}_0$ is defined and an initial state $\Psi_0$ is prepared on the quantum computer.  The time dependent Hamiltonian is
 \begin{equation}\label{RTEham}
 \mathcal{H}(t)=\mathcal{H}_0+\lambda(t)\mathcal{H}_I+\mathcal{C}
 \end{equation}
 with interaction term $\mathcal{H}_I$, constant $\mathcal{C}$, and time dependent coupling constant (or similar form) $\lambda(t)$ at time $t$.  By adiabatically (slowly) increasing the interaction term in time, the wavefunction will eventually arrive at the ground state for the fully interacting Hamiltonian.

However, this solution strategy is known to be extremely slow for quantum chemical systems \cite{poulin2014trotter,lemieux2021resource}.  In order to apply the Hamiltonian, a time evolution operator of the form $\exp(-i\mathcal{H}(t)\delta t)$ must be applied to the wavefunction.  The Trotter-Suzuki decomposition of the time evolution operator must be decomposed into many terms, $\mathcal{O}(N^4)$ to capture the full electron-electron interaction term, although this can be reduced as $N\rightarrow\infty$ to $\mathcal{O}(N^2)$ for the case of local basis functions \cite{baker2020density}. However, since the time step $\delta t$ must be very small depending on the strength of correlation in the system, the resulting number of operations makes the time necessary to solve for even small molecules extremely long. This is true of other classical solution techniques such as Hartree-Fock \cite{gulania2021limitations}. This is expected based on the complexity of solving quantum chemistry systems \cite{schuch2009computational}.

The question of how to obtain excited states is one that has been investigated in recent papers \cite{higgott2019variational,nakanishi2019subspace,parrish2019quantum,mizuta2021deep,kuroiwa2021penalty,greene2021generalized,motta2020determining,jahangiri2020quantum,ollitrault2020quantum,roggero2020preparation,sim2018quantum,bauman2019quantum}, and a direct solution would provide a means to fully manipulate the wavefunction. If this can be accomplished without the use of time evolutions, then  avoiding the small time step necessary for those methods could be possible.  

In addition to ground state solvers, obtaining excited states is a highly valuable quantity for quantum chemistry systems.  Historically, excited states have been more difficult to obtain. So, if a quantum algorithm could reliably obtain the excitations, then this would represent a major improvement over existing classical techniques \cite{elliott2011perspectives,khoromskaia2015tensor,helgaker2014molecular}.
 
Recently, Lanczos methods have been investigated in the context of solving quantum systems.  One variety of Lanczos on the quantum computer, uses even vectors of a Krylov subspace and imaginary time evolution techniques to obtain the ground state and other quantities of interest \cite{motta2020determining,sun2021quantum}. 
 
 Another recently introduced variety of Lanczos algorithms \cite{baker2021lanczos} uses oblivious amplitude amplification (OAA) \cite{kothari2014efficient,berry2015simulating,childs2017quantum}--an algorithm similar to Grover's search but run on auxiliary qubits \cite{grover1996fast,grover2001schrodinger,brassard2002quantum}--to apply operators directly onto wavefunctions. Since OAA applies operators onto wavefunctions, all the necessary operations for Lanczos can be implemented directly.  The full Lanczos recursion can be implemented to find the ground state or the continued fraction representation of the Green's function \cite{baker2021lanczos}. This second Lanczos technique called quantum Lanczos recursion (QLR) avoids the use of a time evolution operator and therefore bypasses the computational bottleneck in terms of the number of terms in the Trotter-Suzuki decomposition.  
 
 One other advantages of the QLR is that the traditional limitations of Lanczos on the classical computer are entirely circumvented \cite{cahill2000numerical}, and only one  number of wavefunctions.  The only limitation is how accurately the operations can be applied to the wavefunction in practice, and this has recently been investigated on a quantum computer \cite{jamet2021krylov}.
 
This paper uses similar techniques as QLR to show that the traditional Lanczos recursion can be replaced by a block Lanczos routine to resolve several excitations, leading to quantum block Lanczos recursion (QBLR), although ``block" can be replaced by ``banded" or some other name corresponding to a trivial change of gauge of the unitary matrices involved.   While it is trivially demonstrated from QLR that excitations can be found, the performance of block Lanczos will allow for the resolving of degeneracies with greater ease. This was recently demonstrated in tensor network algorithms \cite{bakermultitarget,di2021efficient}.  

The algorithm here in the can also be performed with the preparation of a single initialization of starting wavefunctions.  A full collapse of the eigenstates is avoided, meaning that a wavefunction can be preserved for the next computation.  This is accomplished by use of a state-preserving quantum counting algorithm \cite{baker2021lanczos}, referred to in some works as QMA-sampling \cite{marriott2005quantum,temme2011quantum}.

Some additional discussion on how errors in the Lanczos coefficients will affect the ground state energy and otherwise are also considered. The method of slowly introducing terms into the Hamiltonian is also discussed in the context of Lanczos to show that the OAA algorithm does not need to scale exponentially for limited additional terms.

\section{Excitations from quantum Lanczos recursion}

The Lanczos algorithm from Ref.~\onlinecite{baker2021lanczos} can be used to find excitations of a given model by a simple alternative of coefficients.  In this section, QLR will be reviewed and an extension to finding excited states will be shown.

A Lanczos recursion relation to find subsequent elements of the Krylov subspace, $\{\psi_0,\psi_1,\ldots,\psi_N\}$ is 
\begin{equation}\label{scalarlanczos}
|\psi_{n+1}\rangle=\mathcal{H}|\psi_n\rangle-\alpha_n|\psi_n\rangle-\beta_n|\psi_{n-1}\rangle
\end{equation}
where the resulting Hamiltonian in the basis of the Krylov subspace forms a tridiagonal matrix.  When diagonalized, the ground state is found to a high accuracy even if only a few $n$ are determined. 

\paragraph{Applying operators to wavefunctions.--}
The only ingredient that is necessary for this form of the ground state solution is to ensure that the operators could be applied to the wavefunction.  The method to do this was introduced in Refs.~\onlinecite{berry2014exponential,kothari2014efficient} where an oblivious amplitude amplification (OAA) algorithm \cite{berry2015simulating,childs2017quantum} was used to improve on repeat-until-success methods which have the same goal \cite{guerreschi2019repeat,paetznick2013repeat}. Essentially, when an operator is applied to a wavefunction as in a repeat-until-success strategy, the operation is controlled on some auxiliary qubits that give 0 if the application of an operator is successful and 1 if it is not. OAA searches for a state on some auxiliary qubits corresponding to all 0s.  By searching over the auxiliary qubit states, the algorithm remains oblivious to the wavefunction itself. This essentially guarantees the outcome from the related method of repeat-until-success to apply the operator in one step \cite{guerreschi2019repeat,paetznick2013repeat}. For how to construct the operators, see the discussion in Ref.~\onlinecite{low2019hamiltonian}.

\paragraph{Minimal measurements of the wavefunctions.--}
In order to sample the coefficients $\alpha_n$ and $\beta_n$ without completely measuring (and therefore destroying) the wavefunction, a state-preserving quantum counting algorithm can be used.  The algorithm applies a generic operator $\mathcal{A}$ as in the form
\begin{equation}
\mathcal{A}|\Psi\rangle=p|\Psi\rangle+p^{\perp}|\Psi^{\perp}\rangle
\end{equation}
which signifies a superposition of eigenstates. The operator $\mathcal{A}$ is normalized and represented as a unitary such that it can be represented on the qubits \cite{baker2020density,baker2021lanczos}. The resulting probability $p$ is the expectation value of $\mathcal{A}$.  All states orthogonal to the original state are marked with a $\perp$ symbol. The basis of the superposition chosen here is that of the eigenbasis because this will be the natural basis to pick for quantum phase estimation (QPE) \cite{nielsen2010quantum,low2019hamiltonian,poulin2018quantum}.

The key to finding $p$ is to count the number of transitions from $\Psi$ to $\Psi$ (wavefunction to same wavefunction) after the application of $\mathcal{A}$ by OAA or some other algorithm. To verify that the same state is obtained after this procedure, the energy of the state can be computed with QPE at the start of the algorithm and stored on a register.  After the operator is applied, the energy is then computed on a separate register. The two registers are compared and represented on a single qubit, called a pointer qubit.  The pointer qubit is then measured.  If the energies match, one value is returned and the original wavefunction is recovered.  This is counted as a ``success" in the algorithm.  The ratio of successes to the total times the algorithm is run is $p$.  Once $p$ is found, the expectation value of $\mathcal{A}$ can be determined.

If the pointer qubit demonstrates that the wrong state was recovered, then a recovery procedure is used to find the original state \cite{baker2021lanczos,baker2020density,temme2011quantum,paetznick2013repeat}.  In essence, the unitary of all operations applied onto the wrong wavefunction and the process above is repeated until the correct wavefunction is found \cite{temme2011quantum}.

\paragraph{Operators for Lanczos recursion.--}
To find the Lanczos coefficients, the following relations hold \cite{baker2021lanczos}
\begin{equation}
\alpha_n=\langle\psi_n|\mathcal{H}|\psi_n\rangle\quad\mathrm{and}\quad\beta_n=\langle\psi_{n-1}|\mathcal{H}|\psi_n\rangle
\end{equation}
and therefore provide a means to use state-preserving quantum counting to obtain the coefficients. The map between the Krylov states and the original ground state is crucial but can be summarized as
\begin{equation}
|\psi_n\rangle=\hat G_n|\Psi\rangle
\end{equation}
where some examples of these operators are shown in Ref.~\onlinecite{baker2021lanczos}.  The coefficient $\alpha_n$ can be regarded as applying quantum counting on the state $\Psi$ with the operator $(\hat G_n^\dagger\mathcal{H}\hat G_n)$.  Each operator $\hat G_n$ depends on coefficients from $\{0,1,\ldots,n-1\}$ iterations and therefore the algorithm discovers the coefficients iteratively.  To find $\beta_n$, the operator $(\hat G_{n-1}^\dagger\mathcal{H}\hat G_n)$ is used instead.

\paragraph{Energies from QLR.--}
Once the Lanczos coefficients are obtained, the Hamiltonain matrix in its tridiagonal form is then known.  Diagonalizing this matrix and retaining both the energies and eigenvectors makes the new ground state in terms of the Krylov basis chosen.

Defining an operator $\hat Y^{(g)}$ for an excitation $g\in\mathbb{Z}^+$, the operator for the $g$th excitation would be defined as
\begin{equation}\label{Yops}
\hat Y^{(g)}=\sum_n\gamma^{(g)}_n\hat G_n
\end{equation}
in terms of coefficients $\gamma_n$ found from the diagonalization.  The summation over $n$ is over as many elements in the Krylov basis as kept.

Thus, the QLR algorithm can access ground states ($g=0$) or excited states ($g\geq1$) by simply selecting different coefficients from the diagonalization of $\mathcal{H}$.  At the end of this procedure, the ground state is recovered and can be used without fully re-preparing it.

\subsection{Reduced computational complexity of applying operators}\label{complexOAA}

\paragraph{First strategy: apply interaction incrementally.--}
When applying an operator to a wavefunction, several auxiliary qubits are used. OAA is a Grover search over auxuiliary qubits, so it will scale exponentially, $\mathcal{O}(\sqrt{2^q})$ for $q$ qubits, in terms of the number of auxiliary qubits.  While Grover's algorithm promises a square root speedup over classical search methods, the size of the database on the quantum computer is exponentially large in terms of the number of qubits.  This calls into question whether operators can actually be applied to wavefunctions in the manner prescribed earlier in total, but the particular details of the algorithm will show that this can be done efficiently.  This clearly will not present an issue for finding quantities necessary for density functional theory \cite{baker2020density}, since the required operators are small ($q=2$ for coefficients of the one-body reduced density matrix).  

By applying operators from the full Hamiltonian incrementally in groups of size $D$, the cost can be reduced to $\mathcal{O}(D2^{\frac{\lceil q/D\rceil}2})$. Thus, the OAA is feasible here since the Hamiltonian can be expressed as a linear combination of unitaries, this strategy applies to any system of interest \cite{berry2015simulating}. The minimal size of a group of operators is the same as a single term in the Hamiltonian.  For many-body problems, this will be $q=$ 2 or 4 terms for the quadratic single-body terms or the quartic electron-interaction terms \cite{baker2020density}. 

\paragraph{Second strategy: decrease strength of interaction.--}
Note that the operator applied onto the original wavefunction $\Psi$ does not need to be an operator whose eigenvalues include $\Psi$. This means that starting from some known state $\Psi$ with associated operator $\mathcal{H}_0$, 
\begin{equation}
\mathcal{H}=\mathcal{H}_0+\lambda\mathcal{H}_I
\end{equation}
can be defined as a new Hamiltonian.  This form of the Hamiltonian has a similar form as the the adiabatic evolution in Eq.~\eqref{RTEham}, except that the $\lambda$ coefficient can be much larger than the step size in time evolution.

The interaction term, $\mathcal{H}_I$, can be the full interaction term or one term of that interaction.  By adding the terms incrementally, the final state is closer to the initial state provided and will allow for the algorithm to find the new ground state.  It is crucial that each step of this procedure begins with an eigenvector so that the QPE can be applied correctly in the quantum counting procedure.

\begin{figure}
\includegraphics[width=\columnwidth]{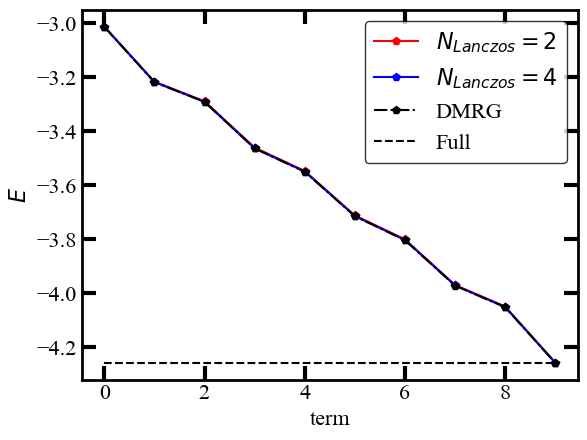}
\begin{picture}(0,0)
\put(-205,35){\includegraphics[height=0.28\columnwidth]{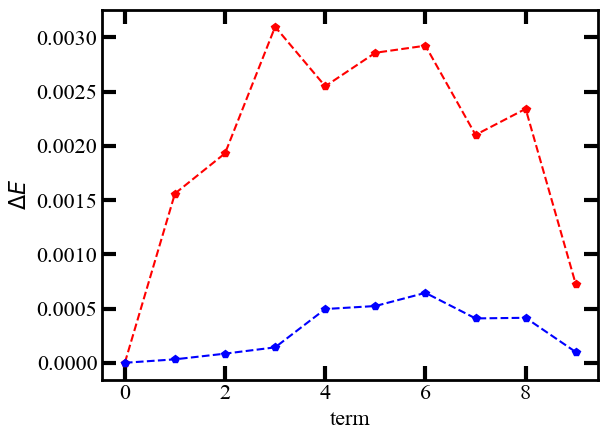}}
\end{picture}
 \put (-175,160) {\Large $\mathbf{J_{xy}=J_z}$}
\caption{\label{incrLanczos} Convergence in energy ($E$) while adding one interaction term $\hat S^z_i\cdot\hat S^z_{i+1}$ to an XY model on a 10 site lattice (9 interaction terms to add).  Two solvers are used at each step. DMRG is shown as a solid blue line and a Lanczos solver is shown as a red dashed line.  The initial wavefunction for the Lanczos algorithm is taken from the previous iteration. A black dashed line shows the energy of the full Heisenberg model. An inset shows the difference between the DMRG and Lanczos solutions. Only one Lanczos iteration was used for each added term. Inset: energy differences from the exact value, $\Delta E$.
}
\end{figure}

\subsection{Demonstration on model systems}\label{demoDMRG}

A full demonstration on real systems by implementing a linear combination of unitaries will be delayed for a future work, but it will be shown here taht Lanczos can be applied iteratively according to the suggestions in the previous paragraphs.

In order to demonstrate that only a limited number of Lanczos steps can be used to solve a model that includes a small number of extra terms from the starting wavefunction's Hamiltonian, a numerical study on a 10-site XXZ model of the form
\begin{equation}
\mathcal{H}=\sum_iJ_{xy}\left(\hat S^x_i\hat S^x_{i+1}+\hat S^y_i\hat S^y_{i+1}\right)+J_z\hat S^z_i\hat S^z_{i+1}
\end{equation}
where $J_{xy}=J_z$ is the Heisenberg Hamiltonian and spin matrices $\mathbf{S}=(\hat S^x,\hat S^y,\hat S^z)=\frac12\boldsymbol{\sigma}$ (with $\hbar=1$) for the vector of Pauli matrices $\boldsymbol{\sigma}$ \cite{townsend2000modern}.  To start, an XY model,
\begin{equation}\label{XYmodel}
\mathcal{H}_\mathrm{XY}=J_{xy}\sum_i\hat S^x_i\hat S^x_{i+1}+\hat S^y_i\hat S^y_{i+1},
\end{equation}
will be solved for the initial wavefunction and a single interaction term of the form $\hat S^z_i\hat S^z_{i+1}$ will be added.  In all, 9 terms will be added to the Hamiltonian. Computations were made with the DMRjulia library using the density matrix renormalization group (DMRG), Lanczos, and exact diagonalization (ED) routines \cite{bakerCJP21,*baker2019m,dmrjulia1,dmrjulia}. 

There are three cases of study here: $J_{xy}=J_z$ (small perturbations), $J_{xy}\ll J_z$ (large perturbations), and when the initial $\psi_0$ is far from the starting state.

\begin{figure}
\includegraphics[width=\columnwidth]{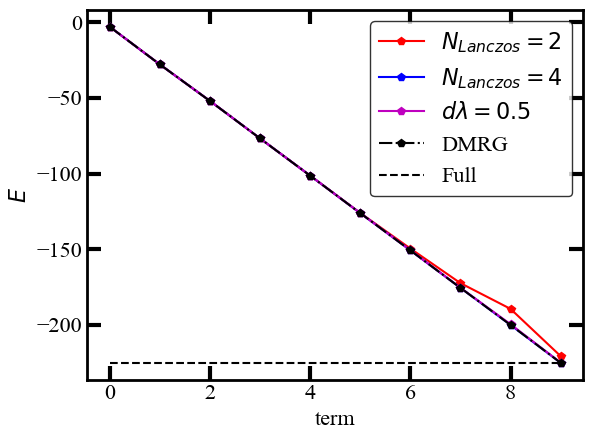}
\begin{picture}(0,0)
\put(-200,32){\includegraphics[height=0.28\columnwidth]{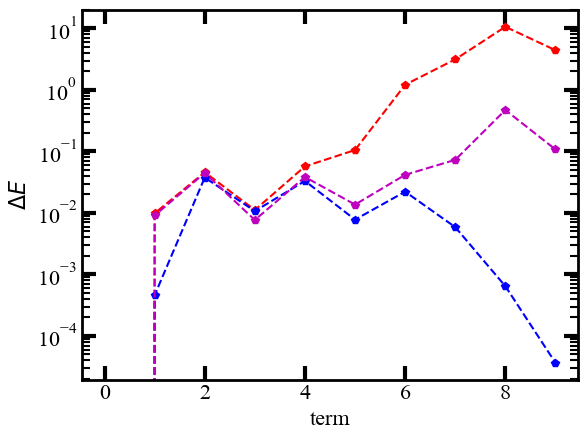}}
\end{picture}
 \put (-170,160) {\Large $\mathbf{J_{xy}\ll J_z}$}
\caption{\label{bigincrLanczos} Convergence in energy ($E$) while adding a large interaction term $\hat J_zS^z_i\cdot\hat S^z_{i+1}$ with $J_{xy}\ll J_z$.  The parameters are the same as FIg.~\ref{incrLanczos}.  By using 4 Lanczos steps instead of 2, the accuracy is greatly improved in this case. Decreasing the incremental interaction strength, $d\lambda$ (here, 0.5 with two rounds of $N_\mathrm{Lanczos}=2$), would require more than double-precision (see text).
}
\end{figure}

\subsubsection{Small perturbations ($J_{xy}=J_z$)}

Figure~\ref{incrLanczos} shows how the energy converges to the energy of the full Heisenberg model with increasing numbers of terms. In each case, using the initial wavefunction provided from the previous iteration, only one Lanczos iteration must be run to obtain a highly accurate energy with the new interaction term included.  The small difference in energies between the DMRG solution and the Lanczos solution is shown in the inset of Fig.~\ref{incrLanczos}. There is also the possibility to add in a partial term by using a small term $d\lambda$ for each of $N$ times such that $Nd\lambda=\lambda$.  This would require more applications of the Lanczos algorithm with new Hamiltonians and may be useful for long-range interactions or other cases.

In this one-dimensional example, the convergence is aided by only adding one term.  In realistic systems, the same strategy of adding interaction terms piecemeal will both allow the Lanczos algorithm to converge quickly and also to reduce the time spent applying the operator with OAA. Still, this should be expected in general due to the rapid convergence of Lanczos techniques.

\subsubsection{Large perturbations ($J_{xy}\ll J_z$)}

In this case, $100J_{xy}=J_z$ to simulate a large perturbation.  For the example here, increasing the number of Lanczos iterations to 4 for each new interaction term added allows for the energy to be obtained to a good accuracy here.  This is a large improvement over using only 2 Lanczos iterations as shown by comparing the curves in Fig.~\ref{bigincrLanczos}.

It is possible to use only 2 Lanczos updates if one adds portions of the interaction term.  In this case, $J_z/2$ is added to the Hamiltonian twice with 2 Lanczos steps each (total 4 Lanczos steps).  The results could be improved if more Lanczos iterations were used here, but this would require greater precision in the coefficients than is available here with a double-precision classical implementation.

Note that tests with coefficients found on the classical computer will be subject to numerical instabilities that will eventually degrade the accuracy of the resulting ground state, so a full solution with this method will generate imprecise answers.  For example, when increasing the number of Lanczos iterations to 10 for each step, this can occur in this example.

\begin{figure}
\includegraphics[width=\columnwidth]{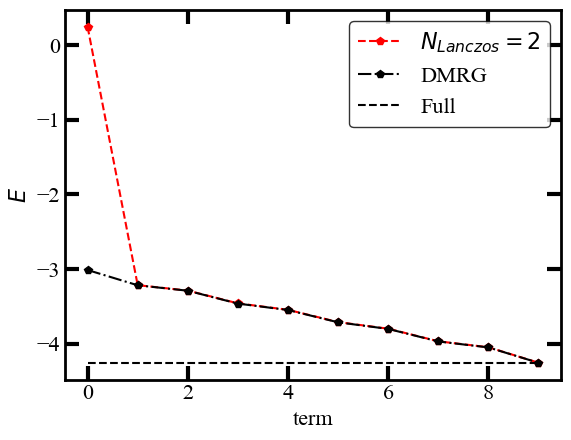}
\begin{picture}(0,0)
\put(-110,55){\includegraphics[height=0.28\columnwidth]{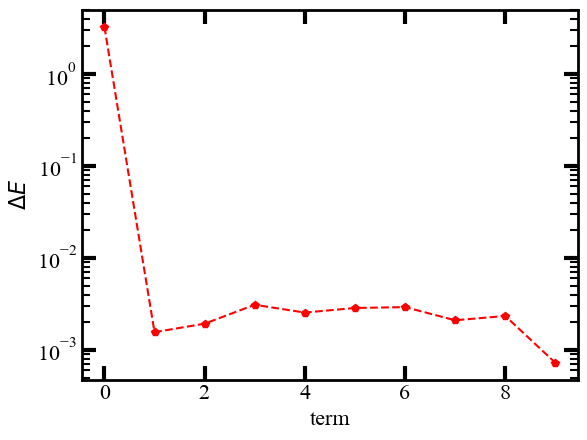}}
\end{picture}
 \put (-190,150) {\Large random $\psi_0$}
\caption{\label{randincrLanczos} Convergence in energy ($E$) similar to Fig.~\ref{incrLanczos} except that a random starting state was used. Tests on larger systems or with more Lanczos recursion steps quickly suffer from precision errors that will not be present on the quantum computer.
}
\end{figure}

\subsubsection{Arbitrary starting wavefunction}

So far, eigenstates for the starting wavefunction were used.  The natural question is whether the starting states could be replaced by some arbitrary state and how the convergence of the algorithm is affected.  When implementing this type of initialization on the quantum computer, the only key is to determine the energy of the wavefunction.  If using QPE to do this, then the starting state must have an associated Hamiltonian. 

For demonstration purposes, the initial state of alternating spins here is
\begin{equation}\label{spinvec}
|\Psi\rangle=|\psi_0\rangle=|\uparrow \uparrow \downarrow \downarrow \downarrow \uparrow \downarrow \downarrow \uparrow \uparrow\rangle
\end{equation}
is taken as the initial state with $J_{xy}=J_z$ in the Hamiltonian.  While the Hamiltonian which has an eigenstate of Eq.~\eqref{spinvec} is not known and would therefore prevent the use of this state in the quantum algorithm, it is instructive to observe the convergence in this case where the initial state is not close to the final problem to solve.

Figure~\ref{randincrLanczos} shows the convergence for this starting wavefunction. The first Lanczos iteration will be the most difficult since it must change the wavefunction the most.  Subsequent iterations converge more easily to the true ground-state just as in the previous cases.

These examples demonstrate that applying operators incrementally is possible and will yield a large cost reduction on the quantum computer.  The examples here were on the ground-state, but higher excited states can be found similar to the discussion around Eq.~\eqref{Yops} once the Lanczos coefficients are known.  In this example, the precision that can be found from the classical simulation is restricted \cite{cahill2000numerical}.  The quantum computer is only limited is not a limitation on the quantum computer assuming perfect application of the operators.

\section{Quantum block Lanczos recursion}

In principle should be able to discover all excitations from QLR as described above \cite{chepiga2017excitation}, but the extension of this method to a block or banded Lanczos algorithm is known to resolve degeneracies to a higher degree and aid convergence generally \cite{bakermultitarget}. Block Lanczos is used in quantum chemistry \cite{bai2000templates} but also in physics, particularly for dynamical mean field theory computations \cite{senechal2008introduction}.  The scalar coefficients of the Lanczos recursion are extended to a matrix of coefficients.


To formulate the problem, consider a set of wavefunctions grouped as a supervectors $\mathbf{\Psi}$ is a vector of $d$ excitations,
\begin{equation}
\mathbf{\Psi}_n=\Big(|\psi_n\rangle_1,|\psi_n\rangle_2,\ldots,|\psi_n\rangle_d\Big)
\end{equation}
and therefore the task is to find matrices $\mathbf{A}$ and $\mathbf{B}$ which are of dimension $d\times d$ block such that Lanczos can be performed.  This means that $d$ registers, each with a wavefunction are also available.

Extending the 3-term recursion from Eq.~\ref{scalarlanczos} to a block or banded (or other) Lanczos scheme with more terms in the recursion would be possible to prepare more than one excitation at a time \cite{senechal2008introduction}.  The expanded Lanczos recursion relation appears as \cite{weikert1996block}
\begin{equation}\label{blockLanczos}
\mathbf{\Psi}_{n+1}\mathbf{B}_{n+1}=\mathcal{H}\mathbf{\Psi}_n-\mathbf{\Psi}_n\mathbf{A}_n-\mathbf{\Psi}_{n-1}\mathbf{B}_{n}^\dagger
\end{equation}
for a vector of wavefunctions $\mathbf{\Psi}$ with matrices
\begin{equation}\label{Asubmat}
\mathbf{A}_n=\mathbf{\Psi}^\dagger_n\mathcal{H}\mathbf{\Psi}_n
\end{equation}
and $\mathbf{B}_n$ defined recursively from
\begin{equation}\label{Bsubmat}
\mathbf{B}_n=\mathbf{\Psi}_{n-1}^\dagger\mathcal{H}\mathbf{\Psi}_n=\mathbf{\Psi}_{n-1}^\dagger\mathcal{H}\mathbf{\Psi}_{n-1}=\mathbf{B}_n^\dagger
\end{equation}
as can be most immediately seen from the tensor network diagrams in Ref.~\onlinecite{bakermultitarget}.    It should be understood that when applying $\mathcal{H}$ onto the wavefunction supervector that the Hamiltonian is applied onto each wavefunction in the vector. The implication of Eqs.~\ref{Asubmat} and \ref{Bsubmat} is that there are simply $d^2$ applications of QLR to obtain QBLR.

Because the coefficients can be obtained by an operator acting on a ground state of an initial Hamiltonian, this implies that the coefficients of both $\mathbf{A}$ and $\mathbf{B}$ can be recovered by performing a state-preserving quantum counting operation on each term of the matrix.  This was also true in the scalar case, and this establishes that the block Lanczos case is in the RWMP category of algorithms thus avoiding repeated wavefunction preparation \cite{baker2020density}.  Each coefficient must be obtained individually: one quantum counting process for each operator.  However,   If operations can be performed in parallel on the quantum computer, then this could improve performance to find all coefficients of a block simultaneously.

On the classical computer, the blocks can be represented as a block diagonal super-matrix representing the Hamiltonian
\begin{equation}\label{blockHam}
\boldsymbol{\mathcal{H}}=\left(\begin{array}{cccccc}
\mathbf{A}_0 & \mathbf{B}_1^\dagger & \mathbf{0} & \cdots & \mathbf{0}\\
\mathbf{B}_1 & \mathbf{A}_1 & \mathbf{B}_2^\dagger & \cdots & \mathbf{0}\\
\mathbf{0} & \mathbf{B}_2 & \mathbf{A}_2 & \ddots & \mathbf{0}\\
\vdots & \vdots & \ddots & \ddots & \mathbf{B}_n^\dagger\\
\mathbf{0} & \mathbf{0} & \mathbf{0} & \mathbf{B}_n & \mathbf{A}_n\\
\end{array}\right)
\end{equation}
which can be diagonalized to find the energies of the excitations.  From the coefficients of the eigenvectors of $\boldsymbol{\mathcal{H}}$ (denoted as $\gamma_{n}$ still here too), the excitation wavefunctions can also be obtained.

In order to understand how best to apply the operators on the quantum computer, the block Lanczos equation can be written as
\begin{align}
\sum_jB^{(n+1)}_{ij}|\psi_{n+1}\rangle_j=&\mathcal{H}|\psi_n\rangle_i-\sum_jA_{ij}^{(n)}|\psi_n\rangle_j\\
&-\sum_j(B^*)_{ij}^{(n)}|\psi_{n-1}\rangle_j\nonumber
\end{align}
for a single row $i$ of the left hand side.  This form makes clear that the $d$ registers containing the $d$ excitations can be acted on with the appropriate operators (sum over $j$) and added with the wavefunction on other registers with the appropriate operators.  Note that the elements of $\mathbf{A}$ and $\mathbf{B}$ are stored classically and therefore the inverse of each matrix can be found.  If this is done, then apply $\mathbf{B}^{-1}_{n+1}$ once the coefficients are found from Eq.~\eqref{Bsubmat}. 

The operators from Eq.~\eqref{Yops} will be denoted as $\mathbf{G}_n$ for a given level and are extended from the scalar Lanczos case as $\hat G_n\rightarrow\mathbf{G}_n$.  Similarly, $\psi\rightarrow\mathbf{\Psi}$ for the wavefunction.  One additional operation is included $\mathbf{G}$, compared with the operators listed in Ref.~\onlinecite{baker2021lanczos}, that of the inverse $\mathbf{B}^{-1}$ operator. Since the operators can be determined on the classical computer, the operators can simply be prepared in a slightly different way.  Again, the inverse is converted to a unitary as is standard for applying operators \cite{baker2021lanczos,baker2020density,berry2015simulating,kothari2014efficient,guerreschi2019repeat}.

In all, the algorithm costs a practical amount of $d^2$ over the scalar Lanczos case. Each element of the matrix equations is simply one application of a set of equations slightly modified from QLR. Note that the effort expended to apply operators onto wavefunctions is not wasted at the end with a single measurement. The state-preserving quantum counting algorithm can be used to obtain the correct expectation values without completely measuring the wavefunction.

\subsection{Algorithm summary}

The following summarizes the previous discussion to demonstrate how to implement block Lanczos on the quantum computer.

\textbf{Quantum block Lanczos recursion: Algorithm for multiple excited-states}

\begin{enumerate}
 \item  A set of starting eigenfunctions $\{|\Xi\rangle\}$ is prepared on $\nu$ registers (one for each wavefunction).
 \item Start a counter at $n=0$.
 \item Construct the operator $\mathbf{G}_n$ representing the appropriate step from Eq.~\eqref{blockLanczos}. This operator spans several excitations.
 \item $\mathbf{G}_n$ is applied onto the current state and quantum counting is used to determine the Lanczos coefficients, given by Eq.~\eqref{Asubmat} and \eqref{Asubmat}.
 \item With the new coefficient(s), the algorithm returns to step 2, increments $n$, and finds the next step of coefficients.
 \item Storing the coefficients classically, the block diagonal Hamiltonian from Eq.~\eqref{blockHam} can be formed and diagonalized.  A set of coefficients $\gamma_n$ can then be used to determine the energies.
\end{enumerate}

The scaling of the method is no worse than the application of the operators at each step, albeit the number of times this must be run for $d$ excitations is $\mathcal{O}(d^2)$.  As pointed out in Sec.~\ref{complexOAA}, the incautious application of OAA will scale exponentially, but to aid convergence and avoid large computations of operations the interaction term can be applied in smaller pieces.  This drastically reduces the number of Lanzcos steps required to obtain the next wavefunction and limits the time necessary to run the OAA algorithm.  It is also possible to use a repeat-until-success strategy here \cite{baker2021lanczos,guerreschi2019repeat,paetznick2013repeat}, but the guaranteed result from application of OAA probably will save time in the long run.

It is not known beforehand the coefficients of the next step, so each step must be performed iteratively.  Once the coefficients are obtained with the quantum counting process, then they can be stored classically and used without obtaining them again, although the $d^2$ coefficients of a given block can be parallelized. Note also that the algorithm can be restarted at any time since the coefficients are stored classically.  

The ground state energy can be checked when diagonalizing Eq.~\eqref{blockHam} on the classical computer.  If the ground state energy is converged within the accuracy needed for QPE, then there are sufficient Lanczos steps run to find an accurate ground state energy.

\subsection{Error analysis}

In this section, a careful analysis of the relevant parameters of the system will show how uncertainty in the Lanczos coefficients will influence the resulting energies.

\begin{figure}
\includegraphics[width=\columnwidth]{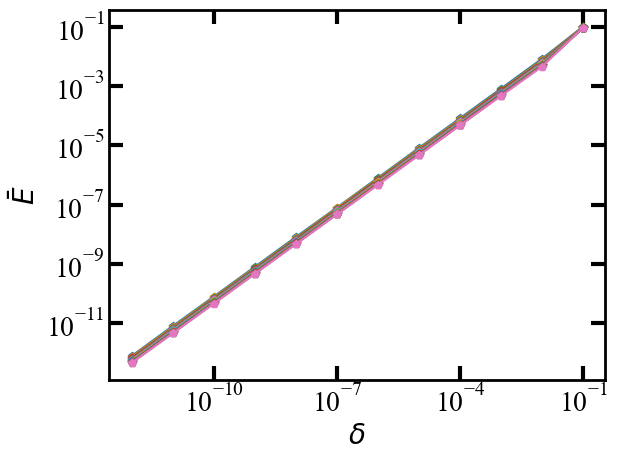}
\caption{Mean average error for a system of block size $d=20$ for various numbers of iterations $\{4,5,6,\ldots,20\}$. All lines are nearly the same in magnitude. The general linear trend is mostly unchanging with the input parameters. \label{mae_energies}}
\end{figure}

To investigate the effects of noise on the resulting energies, a block Hamiltonian of a particular size $a$ is created for $b$ such blocks. When $a=1$, the QLR algorithm is used.  When $a>1$, QBLR is represented.  A set of matrices are generated with values between $[0,1]$ to mimic the values on the operators for the quantum computation. A noise term selected from a Gaussian distribution of width $\eta$ is selected for each element. 

The energies with noise $E$ are computed with respect to the true energies without noise $E_0$ to find the mean absolute error (MAE) $\bar E$ defined as
\begin{equation}
\bar E=\frac1M\sum_{i=1}^M\left|E^{(i)}-E^{(i)}_0\right|
\end{equation}
for a number of eigenvalues $M$.

The MAE is shown in Fig.~\ref{mae_energies}. As expected, as the noise decreases in each of the parameters, the accuracy of the energy eigenstates increases.  In both cases, the error decreases linearly with the noise applied to each term.  This is a straightforward but useful result.   If the quantum counting obtains coefficients to a precision $\delta$, then the resulting energy is also obtained to an error of roughly $\delta$.  The change in this trend for other parameters that could have been picked (larger block sizes, more blocks, etc.) is very stable.

What this implies is that for a QPE to obtain the correct eigenvalue, the Lanczos coefficients must be sufficiently accurate ({\it i.e.}, $\delta$ is a threshold for how accurate the coefficients will be). This is a crucial aspect to know how much precision must be obtained from the quantum counting algorithm.

\subsection{Discussion}

The advantages of many excitations being discovered is chiefly one of stability.  The ability to resolve degeneracies and generate orthogonal wavefunctions that is valuable in a variety of contexts and now available for the quantum computer.  That this method is rapidly convergent \cite{ye1991convergence,cullum1996arnoldi}, resolves degeneracies, and avoids repeated wavefunction preparation means that this method could be a valuable alternative to time evolution methods.

Note that unlike in Ref.~\onlinecite{bakermultitarget}, a singular value, QR, or LQ decomposition \cite{press1992numerical} was not used.  This is primarily because it was not necessary here, but it has several advantages.  First, many of these algorithms on the quantum computer require a quantum memory, which may be far off in terms of development \cite{chia2020sampling,tang2018quantum}, particularly with the growing consensus that probably approximately correct (PAC) learning on the quantum computer is not likely to be the best way forward for quantum machine learning \cite{baker2020density,huang2021information}. Although, expanding the definition has been suggested to lead to a logarithmic dependence on the number of measurements for tomography in terms of the total Hilbert space size \cite{aaronson2019shadow,huang2021provably}, finding the initial state for the tomography could be accomplished with QLR or QBLR.  

The goal of QBLR is to retain the accuracy of the wavefunction, which is controlled directly through the gate fidelities and other quantities on the quantum computer.  The quantum advantage sought here is for an exponential reduction in memory, which is a property that quantum computers exhibit under the extended Church-Turing thesis which asks if an algorithm can be computed with significantly less memory resources over classical computing \cite{aaronson2011computational}.  Lanczos techniques do have this property when applied on a quantum computer theoretically if perfect application of gates is assumed and with sufficient time to sample the coefficients.  The typical errors on a classical computer \cite{cahill2000numerical,paige1976error,bai1994error,chepiga2017excitation,aaronson2011computational} do not appear since values are retained to quantum precision.

The opportunity to obtain ground states from QLR places added emphasis on the development of state-preserving quantum counting and methods of applying operators to wavefunctions such as the linear combination of unitaries (and included subroutines such as OAA) to obtain both the Green's function and ground-state. 

The possibility to discover ground states or excited states deserve more consideration since these methods are free of time steps and Trotter-Suzuki decompositions, a common feature in many other proposals. The Lanczos algorithms also do not need any notion of locality to aid convergence as would be expected for a matrix product state's relationship to local Hamiltonians \cite{dmrjulia1,verstraete2006matrix,hastings2004locality}, particularly for time-evolving block decimation (TEBD) \cite{bakerCJP21}.  The only instance where locality plays any role is in the writing of the application of the operators to the wavefunction, which could play a role in near-term implementation. It is not clear how many Lanczos coefficients would need to be obtained to find accurate ground states from this method for phase-estimation to give accurate results. Further discussion would require specialization to a specific problem and is deferred to a future study.

Note that extensions to a fitting function for the remaining coefficients is possible in principle \cite{viswanath2008recursion}.

\section{Non-Hermitian operators}

The previous discussion was for the block Lanczos algorithm that is formulated for the problem of a Hermitian operator. The extension of the above methods to non-Hermitian operators is possible without introducing too much additional computational cost.  Instead, only one additional set of matrices (or in the case of scalar Lanczos, one additional set of coefficients) should be found.  The procedure will be define for the block-matrix case, but it can be reduced to the scalar case when the block size is of size 1.

Given a non-Hermitian operator $\tilde{\mathcal{H}}$ with eigenvalue $E$, a set of two eigenvalues can be defined.  One is known as the left-eigenvectors (a transpose is used even if there are complex valued entries) \cite{bai1994error}
\begin{equation}
\mathbf\Psi_L^T\tilde{\mathcal{H}}=\mathbf\Psi_L^T E
\end{equation}
 and the other is for the right-eigenvectors,
\begin{equation}
\tilde{\mathcal{H}}\mathbf\Psi_R=E\mathbf\Psi_R
\end{equation}
where block matrices $\mathbf{\Psi}_L$ and $\mathbf{\Psi}_R$ have the orthogonality relation \cite{bai1994error}
\begin{equation}
\mathbf{\Psi}_L^T\mathbf{\Psi}_R=\mathbb{I}_{d\times d}
\end{equation}
with an identity matrix $\mathbb{I}_{d\times d}$ of dimension $d\times d$.

The block operator represented analogous to Eq.~\eqref{blockHam} appears in this case as
\begin{equation}
\mathbf{\check T}_n=\left(\begin{array}{cccccc}
\mathbf{A}_0 & \mathbf{C}_1 & \mathbf{0} & \cdots & \mathbf{0}\\
\mathbf{B}_1 & \mathbf{A}_1 & \mathbf{C}_2 & \cdots & \mathbf{0}\\
\mathbf{0} & \mathbf{B}_2 & \mathbf{A}_2 & \ddots & \mathbf{0}\\
\vdots & \vdots & \ddots & \ddots & \mathbf{C}_n\\
\mathbf{0} & \mathbf{0} & \mathbf{0} & \mathbf{B}_n & \mathbf{A}_n\\
\end{array}\right).
\end{equation}
There are two recursion relations for each of the left and right eigenvectors.  They take the form of \cite{bai1999able,gruning2011implementation}
\begin{align}
\vec{\mathbf{\Psi}}_{L,(n)}^T\tilde{\mathcal{H}}&=\mathbf{\check T}_n\vec{\mathbf{\Psi}}_{L,(n)}^T+\mathcal{I}_n\mathbf{C}_{n+1}\mathbf{\Psi}^T_{L,(n+1)}\\
\tilde{\mathcal{H}}\vec{\mathbf{\Psi}}_R&=\vec{\mathbf{\Psi}}_R\mathbf{\check T}_n+\mathbf{\Psi}_{R,(n+1)}\mathbf{B}_{n+1}\mathcal{I}_n^T
\end{align}
where
\begin{align}
\vec{\mathbf{\Psi}}_{R,(n)}&=\left(\begin{array}{ccccc}
\mathbf{\Psi}_R^{(0)}, & \mathbf{\Psi}_R^{(1)}, & \ldots, & \mathbf{\Psi}_R^{(n)}
\end{array}\right),\\
\vec{\mathbf{\Psi}}_{L,(n)}&=\left(\begin{array}{ccccc}
\mathbf{\Psi}_L^{(0)}, & \mathbf{\Psi}_L^{(1)}, & \ldots, & \mathbf{\Psi}_L^{(n)}
\end{array}\right),\\
\mathrm{and}\quad\mathcal{I}_n&=\left(\begin{array}{cccc}
\mathbf{0}_{d\times d},&\mathbf{0}_{d\times d},&\ldots,&\mathbb{I}_{d\times d}
\end{array}\right).
\end{align}
The final term is an identity matrix to ensure that the dimension of the resulting matrix is consistent with all terms in the recursion. Only the last block of the matrix is non-zero where blocks of zero matrices, $\mathbf{0}_{d\times d}$.

Writing out the explicit terms of the recursion relation in the style of the $\mathbf{G}$ operators from earlier is lengthy, but a single operator can be written to be applied on the starting wavefunction just as for QLR \cite{baker2021lanczos}.  The only additional cost is the determination of an extra block matrix, $\mathbf{C}$, and double the number of registers to represent both the left- and right-eigenvectors.

The extension of the block Lanczos algorithm to the non-Hermitian case has therefore only introduced an extra matrix $\mathbf{C}_n$ which must also be found with the other two sets of matrices.  However, this does not increase the overall computational cost of the algorithm presented here.  The same forms for the operators connecting the original wavefunction provided to the algorithm can be derived similarly as for the Hermitian case.

\section{Conclusion}

Lanczos recursion methods on the quantum computer were extended to solve for many excitations simultaneously. This uses a block Lanczos technique that is good at resolving degeneracies in quantum states.  This comes at only a cost of the number of excitations squared sought on the quantum computer.  The use of the quantum counting algorithm here allows for the wavefunction to not be collapsed at each step, cutting out a major cost of many other algorithms. Further, interaction terms can be applied in small groups to aid convergence and keep the process of applying those operators to the wavefunction less than exponentially long. The error of diagonalizing the Hamiltonian  in the Krylov basis was demonstrated to scale linearly with the noise of the coefficients. The method can also be applied onto non-Hermitian operators with a moderate additional cost.  Due to the feasible cost of this algorithm and the rapid convergence of Lanczos techniques, it is expected that quantum block Lanczos recursion could be an alternative to existing methods to find the ground and excited states.

\section{Acknowledgements}

The author thanks Alexandre Foley, David Sénéchal, Anirban Chowdhury, and David Poulin for enlightening conversations and to Rex Godby for hosting. 

The author is grateful to the US-UK Fulbright Commission for financial support under the Fulbright U.S. Scholarship programme as hosted by the University of York.  This research was undertaken in part thanks to funding from the Bureau of Education and Cultural Affairs from the United States Department of State.

This project was undertaken on the Viking Cluster, which is a high performance compute facility provided by the University of York. T.E.B. is grateful for computational support from the University of York High Performance Computing service, Viking and the Research Computing team.

\bibliography{QBLR,TEB_papers,TEB_books}

\end{document}